\documentclass[manuscript]{acmart}

\AtBeginDocument{%
 \providecommand\BibTeX{{%
 Bib\TeX}}}

\setcopyright{acmcopyright}
\copyrightyear{2026}
\acmYear{2026}
\acmDOI{XXXXXXX.XXXXXXX}

\acmConference[DIS '26]{Designing Interactive Systems 2026}{June 13--17,
  2026}{Singapore, Singapore}

\usepackage{graphicx}
\usepackage{makecell}

\usepackage[ruled]{algorithm2e} 
\usepackage{natbib}
\usepackage{enumitem}

\usepackage{siunitx}
\usepackage{multirow}
\usepackage{subcaption}
\usepackage[justification=centering]{caption}
\usepackage{tabularx}

\usepackage{color, colortbl}
\usepackage{ragged2e}

\RequirePackage[normalem]{ulem} 
\RequirePackage{color}\definecolor{RED}{rgb}{1,0,0}\definecolor{BLUE}{rgb}{0,0,1} 

\AtBeginDocument{%
  \providecommand\BibTeX{{%
    \normalfont B\kern-0.5em{\scshape i\kern-0.25em b}\kern-0.8em\TeX}}}

\begin{document}

\title[Designing Conversations with the Dead]{Designing Conversations with the Dead: How People Engage with Generative Ghosts}

\author{Jack Manning}
\email{jack.manning@colorado.edu}
\affiliation{%
 \institution{University of Colorado Boulder}
 \city{Boulder}
 \state{Colorado}
 \country{USA}
}

\author{Daniel Sullivan}
\email{daniel.sullivan@colorado.edu}
\affiliation{%
 \institution{University of Colorado Boulder}
 \city{Boulder}
 \state{Colorado}
 \country{USA}
}

\author{Dylan Thomas Doyle}
\email{dylan.doyle@colorado.edu}
\affiliation{%
 \institution{University of Colorado Boulder}
 \city{Boulder}
 \state{Colorado}
 \country{USA}
}

\author{Anthony T. Pinter}
\email{anthony.pinter@colorado.edu}
\affiliation{%
 \institution{University of Colorado Boulder}
 \city{Boulder}
 \state{Colorado}
 \country{USA}
}

\author{Jed R. Brubaker}
\email{jed.brubaker@colorado.edu}
\affiliation{%
 \institution{University of Colorado Boulder}
 \city{Boulder}
 \state{Colorado}
 \country{USA}
}

\renewcommand{\shortauthors}{Manning et al.}

\begin{abstract}
We examine how people experience two choices in the design of generative ghosts, AI systems that are trained on data of the dead: representation, where an AI speaks about a deceased person in the third person, and reincarnation, where the AI speaks as the deceased in the first person. Through a qualitative user study with 16 participants, we explore how each shaped authenticity, affect, and risk. Reincarnation was preferred for its immediacy, but participants shared fears of over-reliance. Representation was preferred for engaging with memory over conversational presence, though participants often ignored this distinction, engaging in dialogue despite third-person framing. Across both modes, participants privileged affective resonance over factual fidelity. We conclude by showing how factors such as tone, language, and conversational rhythm -- factors unique to the user's memory of the deceased -- shape interactions with generative ghosts, and argue that those interactions are always collaborative.
\end{abstract}

\ccsdesc[500]{Human-centered computing~Empirical studies in HCI}\ccsdesc[300]{Human-centered computing~User studies}
\ccsdesc[500]{Computing methodologies~Artificial intelligence}
\keywords{generative AI, generative ghosts, human-centered AI, digital legacy, chatbots}

\maketitle

\section{Introduction}
Generative ghosts, AI systems trained on data about the deceased, are no longer speculative \cite{MorrisBrubaker}. Companies are already experimenting with services that promise ongoing connections with lost loved ones \cite{hollanek2024griefbots}, raising new questions about how people relate to technologies that blur memory, presence, and simulation. While technical capabilities continue to advance, how these systems shape experiences of bereavement remains underexplored \cite{manevich2025artificial, MorrisBrubaker,krueger2022communing}. Research has examined how technology design choices shape comfort in end-of-life contexts \cite{Fergusonetal}, but the impact of generative ghost interactions specifically requires further investigation. Additionally, scholars have raised concerns about how generative ghost systems might affect the psychological well-being, dignity, and autonomy of bereaved and grieving users \cite{lindemann2022ethics}, as well as issues of consent and the potential for psychological distress \cite{hollanek2024griefbots}. While this research area raises broad questions (\textit{e.g.}, ownership, family dynamics, accurate depictions), these questions are predicated on the design of these systems. Yet little empirical work has examined how people actually experience and interpret generative ghosts at the interaction level, particularly when interacting with systems grounded in their own personal relationships with deceased individuals.


In this paper, we focus on a central choice in the design of generative ghosts: the point of view from which the ghost speaks. Morris and Brubaker describe this dimension as \textit{anthropomorphism}: ``whether a generative ghost presents itself as a \textit{reincarnation} of the deceased individual or as a \textit{representation} of that individual.'' \cite{MorrisBrubaker}. In the context of chatbots and LLMs, these choices are most evident in 
linguistic choices, such as the point of view the chatbot takes when interacting with a person. In representation, the system describes the deceased in the third person (\textit{e.g.}, ``Jenny wants you to know that she misses you'', or ``Jenny's favorite flower was the tulip.''). In reincarnation, the system speaks as the deceased in the first person (\textit{e.g.}, ``I miss you'', or ``I love tulips''). 


To explore reincarnation vs. representation in generative ghosts, we conducted a user study with 16 participants using an AI-assisted Wizard of Oz setup. Participants interacted with two different modes of a prototype that was provided details about a deceased loved one. 
After engaging in separate chat-based conversations with each mode, participants reflected on their experiences, comparing the two versions. Asking participants to interact with both versions allowed us to examine how they oriented to each, how they made sense of those experiences, and their preferences between the two points of view. 



We found that differences between point of view were less impactful than we anticipated, with some participants actively ignoring the distinction. Meanwhile, our analysis of interactions and participant feedback highlighted how each mode produced different forms of closeness, the importance of emotional affect over factual accuracy, and how people evaluate the benefits and risks of generative ghosts for themselves vs. others.

Building on these findings, in our discussion, we identify additional design attributes beyond narrative point of view (i.e., \textit{anthropomorphism}) to guide designers of generative ghosts: linguistic familiarity, emotional tone, and style and rhythm. Additionally, we move beyond the design of the ghost to highlight the centrality of the interpersonal relationship as a lens through which ghosts are evaluated. We conclude by proposing priority areas for future work to examine the affordances and limitations of personalization in generative ghost design.


\section{Related Work}

To contextualize and motivate our work, we first provide an overview of digital legacy scholarship in HCI. We then discuss more recent work on generative ghosts, deadbots, and the importance of anthropomorphism.

\subsection{Digital Legacy}
For the last two decades, HCI designers have shown increasing interest in digital legacy, broadly conceptualized as the data, accounts, and other digital material passed down by the deceased to recipients such as bereaved loved ones \cite{doyle2023digital}. HCI research on digital legacy has primarily examined platform support at the end of life, focusing on how digital accounts, data, and identities are managed around death \cite{brubaker_legacy_2016,brubaker_stewarding_2014,brubaker2013beyond,doyle2023digital,gulotta_legacy_2014}. As summarized by Doyle and Brubaker \cite{doyle2023digital}, this literature addresses digital legacies as expressions of identity, how people engage with them over time, and how they are ultimately put to rest, with most work centered on social media platforms.

A substantial portion of prior work focuses on memorialization, examining how post-mortem accounts and digital artifacts are repurposed as online memorials and communal spaces for the bereaved \cite{brubaker2011we,moncur_emergent_2014,arnold2017death,brubaker_legacy_2016,gulotta_legacy_2014}. Research in this area highlights design challenges related to shared audiences, ongoing communication, and the preservation of the decedent’s digital identity over time, and proposes stewardship-based models -- such as Facebook’s Legacy Contact -- to manage these complexities \cite{brubaker_stewarding_2014,brubaker_legacy_2016}. Other work shows that bereaved users often seek access to accounts and data for archiving or transfer, but encounter barriers due to platforms’ single-user assumptions and security mechanisms \cite{baumer2017post,adams_whats_2014,holt_personal_2021}. More recent studies have begun to consider multi-platform digital legacy management, including work on privacy, security, and digital data planning across systems \cite{holt_personal_2021,holt2024post,doyle2024so}. Yet a recent audit of post-mortem support across popular online platforms found that most platforms still fail to provide any meaningful end-of-life support \cite{doyle2025assessing}. 

Recently, digital legacy literature has begun to consider the implications of generative AI on memorialization, post-mortem support, grief support, and digital representation of the identity of the deceased \cite{brubaker2024ai}. In contrast to earlier digital legacy research, which focused on more `static’ representations of the deceased, such as social media profiles, media archives, and online memorial pages, a new generation of AI-based systems offers novel functionality: the ability to generate new content.

\subsection{Generative Ghosts \& Anthropomorphism}

Morris and Brubaker define ''generative ghosts'' as agentic AI systems trained on data of deceased individuals that can generate novel content, sometimes with autonomous action-taking capabilities \cite{MorrisBrubaker}. They map a design space that helps differentiate possible AI systems along seven key dimensions: provenance (first-party vs. third-party creation/authorization), deployment timeline (pre-mortem ''clone'' that later transitions vs. post-mortem-only), anthropomorphism (framed as \textit{reincarnation} vs \textit{representation}), multiplicity (one ghost vs multiple versions for different audiences), cutoff date (static vs evolving with time and new information), embodiment (virtual chatbot vs richer/physical embodiments), and representee type (human vs non-human, \textit{e.g.}, pets). While the generative ghost design space was created to describe a wide range of future systems, the most common example today of a generative ghost is a deadbot.


Deadbots, often called griefbots, are AI-driven chatbots that simulate how a deceased individual might speak, with the aim of enabling posthumous interaction or supporting remembrance \cite{hollanek2024griefbots}. The idea has appeared both in research prototypes and in real-world systems: for example, widely reported cases describe people using tools like Project December\footnote{https://projectdecember.net/} to converse with an AI model trained on a deceased loved one’s texts \cite{Fagone_2021}, while commercial products such as HereAfter AI\footnote{http://hereafter.ai/} market ``interactive memory'' experiences that let family members ask questions and hear answers in the person’s recorded voice. Alongside these developments, scholars have raised questions about how to balance the potential benefits and risks associated with these new tools \cite{lei2025ai, MorrisBrubaker}.

Generative ghosts (including deadbots) may be beneficial for the bereaved as they can support continuing bonds \cite{Klass1996ContinuingB} and meaning-making after a loss \cite{xygkou2023conversation}. As with other digital legacy technology, they may provide comfort in specific moments of grief if designed with care, offering a structured way to revisit memories, ask questions, or say things left unsaid \cite{MorrisBrubaker,krueger2022communing,manevich2025artificial}. Moreover, as highlighted by Lei et al. \cite{lei2025ai}, people may feel comforted by or even feel hopeful at the idea of their own generative ghost, ``ensur[ing] that they are remembered, whether by family, friends, or perhaps the world at large.''


At the same time, critics warn that these systems can introduce serious risks: they may intensify distress or complicate grieving, encourage over-reliance on an artificial `presence,' and create opportunities for manipulation or unwanted `digital hauntings' absent strong safeguards \cite{hollanek2024griefbots}. Manevich and Aluma \cite{manevich2025artificial}, taking a middle ground, argue for the importance of regulation, theory, research, and careful clinical application in order to ensure benefits and prevent harm.
It is worth noting that in all these cases, concerns are strongly connected to Morris and Brubaker's feature \textit{anthropomorphism}, focusing on systems that act as the deceased loved one. 



While recent conceptual scholarship, such as the generative ghost design space, is a helpful starting point, the challenge of translating these frameworks into actual ghost design remains an urgent gap. Survey research has found that older individuals are more likely to have ethics-based resistance to using these systems \cite{fu2025ethical}, with emotional support, relief from negative feelings, and escapism as the key drivers of acceptance \cite{fan2025adoption}. Yet it is striking that, to the best of our knowledge, no user studies have been conducted examining hands-on experiences of users with generative ghosts.

When designing generative ghosts, anthropomorphism most closely aligns with how scholars discuss deadbots, including both their potential benefits and their risks. For this reason, we focus on anthropomorphism in this paper. 
Beyond generative ghosts, studies of chatbots, relational agents, and avatars show that design decisions about voice, perspective, and embodiment shape whether systems feel authentic or trustworthy \cite{Bickmoreetal, go2019humanizing}. In these contexts, anthropomorphism has often been treated as a design lever: something that can be dialed up or down through choices about persona and style. 

Building on the work above, we extend digital legacy scholarship by examining how emerging AI technologies shape experiences of memory, identity, and loss. We also advance the growing literature on generative ghosts by offering empirical evidence about end users’ experiences interacting with these systems.

\section{Methods}

\begin{table}[t]
\caption{Overview of participants including age, gender, relationship to the deceased, and time since the deceased passed}
\label{table:demos}
\begin{tabular}{lccll}
\toprule
Participant & Age & Gender & Relationship to Deceased & Time Since Passing \\
\midrule
P01 & 28 & Woman    & Brother                  & $<$1 year \\
P02 & 34 & Man      & Father                   & $>$20 years \\
P03 & 30 & Man      & Grandfather              & 3 years \\
P04 & 32 & Woman    & Grandmother              & 5 years \\
P05 & 32 & Man      & Grandfather              & $>$20 years \\
P06 & 22 & Man      & Grandfather              & 10 years \\
P07 & 25 & Man      & Grandfather              & 3 years \\
P08 & 22 & Woman    & Mother                   & 5 years \\
P09 & 22 & Non-binary & Friend                  & 2 years \\
P10 & 23 & Man      & Stepfather               & 9 years \\
P11 & 32 & Woman    & Grandmother              & $<$1 year \\
P12 & 34 & Man      & Grandmother              & 1 year \\
P13 & 38 & Woman    & Grandmother              & 5 years \\
P14 & 32 & Nonbinary & Family Friend (Uncle)   & 20 years \\
P15 & 50 & Woman    & Grandmother              & 3 years \\
P16 & 44 & Woman    & Grandmother              & 6 years \\
\bottomrule
\end{tabular}
\end{table}
We conducted a qualitative user study to compare two implementations of generative ghosts: representation, in which the system described the deceased in the third person, and reincarnation, in which the system spoke as the deceased in the first person. The study followed a within-subjects design \cite{greenwald1976withinsubjects}, exposing each participant to both conditions in randomized order. Our data collection and analysis were shaped by our positionality as researchers situated within Western, Anglo-American contexts. Additionally, all researchers have experienced the loss of a close loved one, which informed how we approached and interpreted this work. All aspects of the study were approved by the ethics board at the University of Colorado Boulder. Recognizing the emotional nature of the topics being discussed, we implemented procedures to help minimize the emotional risk that participants might face from participating. We were sensitive to signs of distress \cite{brubaker2018protocol}, and resources for counseling and mental health services were made available as part of the study materials.

\subsection{Participants}
We recruited 16 participants through social media and snowball sampling \cite{biernacki1981snowball}, interviewing participants until we reached saturation on attitudes and comparisons between \textit{representation} and \textit{reincarnation} (as defined by \cite{MorrisBrubaker}). All participants were at least 18 years old, had experienced the death of a close relative or friend, and were willing to engage with AI systems based on that person. Participants ranged in age from 22-50. Each participant selected one deceased individual as the individual to interact with. Before the session, participants completed a short intake survey that gathered their demographics and a brief description of the chosen person, including their relationship, personality traits, and communication style. We report participant demographics in Table \ref{table:demos}.

\subsection{Procedures}
Sessions were conducted remotely over Zoom and lasted between 45 and 90 minutes (60-minute average). Two members of the research team collaboratively ran each session: one acted as the `facilitator' while the other acted as the `operator' running the ghosts behind the scenes. The same two members of the research team also led the analysis process.

At the beginning of each session, the facilitator started by explaining the purpose of the study, emphasizing the simulated nature of the system, and obtained informed consent. The facilitator also asked participants about their relationship with the deceased and their emotional readiness to engage with a generative ghost, emphasizing participants' right to withdraw at any time. To avoid ethical concerns associated with deception studies, the operator was present and introduced at the beginning of the study, but turned off their camera before starting either condition. The facilitator remained available throughout.

The main portion of the session consisted of the participants interacting with two generative ghosts based on the information they had provided in their intake survey, one at a time. Participants interacted with each ghost for approximately 20 minutes. Our design allowed each participant to compare how both modes captured a single relationship.
We varied the order in which participants encountered the two to counteract how familiarity with the first mode influenced how the second was evaluated \cite{Keppel2004}.
After interacting with both conditions, participants took part in a semi-structured interview in which they reflected on their experiences, compared the two conditions, and shared their broader perspectives on the role of generative ghosts. We include the interview protocol in the supplementary material that accompanies this paper.


\subsection{Implementation}
We used an AI-assisted Wizard-of-Oz (WoZ) \cite{woz} setup to generate generative ghost responses. We used a WoZ setup as they are particularly well-suited to connect speculative or emerging technologies to design insights \cite{Kim, Muerer, Anderson}. 

We used Zoom's built-in chat functionality as the interface for participants to interact with the ghosts. The facilitator remained on camera throughout the session, while the operator turned off their camera and changed their display name to ``operator'' to maintain the sense of interacting with a system rather than a person. Each condition began with an initial message tailored to the conversation topic the participant had mentioned at the start of the interview (\textit{e.g.}, asking a grandparent what their college experience was like), establishing a natural starting point for the exchange. In representation, the system introduced itself as a third-person narrator, akin to a family archivist, speaking about the deceased, while in reincarnation it spoke directly as the person in the first person. Beyond the textual differences, there were no visual distinctions between the two conditions.

The operator used GPT-4 to generate responses, acting as an intermediary between the Zoom chat and a ChatGPT\footnote{https://chatgpt.com/} interface. The operator pre-seeded the AI conversation with participant-provided descriptions of their loved one from the intake survey and the beginning of the session as grounding material for each chat condition. For the representation condition, the system prompt we used also instructed the model to describe the deceased in the third person, speaking about them as a knowledgeable observer. For the reincarnation condition, the prompt instructed the model to respond as the deceased person in the first person, drawing on the provided personality traits and communication style to embody their voice. 


The speed of the conversation was slow, with responses taking approximately 60-90 seconds on average. This was due to the manual process of the operator copying the message, prompting the model, reviewing and editing the output, and entering it into the chat. While a noteworthy limitation, this quality control ensured the conversation remained on topic while also acting as a precaution against any potentially harmful AI-generated messages. While we had anticipated the operator needing to edit the responses extensively, in practice, responses were only lightly edited. Moreover, the operator regenerated responses very sparingly, only when outputs felt disconnected from the current flow of conversation. 


\subsection{Dataset and Analysis}
Our data consisted of Zoom chat logs from both modes (capturing the real-time exchanges between participants and ghosts) and interview transcripts. The combination of chat logs and interview transcripts provided both interactional data and reflective commentary, allowing us to analyze how participants responded in the moment and how they interpreted the experience afterward \cite{Lazar2017interviews}. 


To analyze our data, we conducted a thematic analysis, informed by Braun and Clark \cite{braun2012thematic}, in conjunction with Saldana's approach to coding \cite{saldana2011fundamentals}. We approached analysis with an orientation towards comparing participant attitudes and preferences about each mode. Analysis took place concurrently with data collection, with the two members of the research team who were involved directly in data collection noting potential factors and memoing about their impact for each participant. 

As our data collection finished, these trends formed the basis of our analytic codebook, allowing us to code data to describe these trends. 
After completing data collection (in accordance with established norms around saturation, where we observed no new trends emerging in the chat logs or interview data \cite{corbin2008basics}), we then returned to our  factors, organizing them into categories and further memos, grounding each in the data. At times, it was helpful to perform a careful analysis of the chat logs, particularly when participants noted word choices and specific exchanges that were impactful. 



Finally, we performed an axial analysis on the memos \cite{corbin2008basics} considering the myriad desires, use-cases, and interactions that participants had with ghosts, and how they made sense of those wants and experiences after the interactions had concluded. Our axial analysis highlighted tensions across both modes and participants -- particularly around themes such as connection, factuality, and implications of such systems. This analysis provided the basis of our four main findings, which we detail next in the Findings section.



\section{Findings}
Our analysis revealed how participants evaluated \textit{reincarnation} and \textit{representation} as two related ways of connecting with those they had lost. They described reincarnation as more vivid and emotionally charged, offering a sense of renewed presence that could comfort as well as unsettle. In contrast, representation created reflective distance that some found stabilizing, but others experienced as detached. The majority of participants discussed the emotional benefits of closeness alongside the risks of growing too attached, wondering how continued use might make it more difficult to gain balance in their grief. While both designs were intended to foster connection, reincarnation more frequently exposed the tension between comfort and dependency. In this section, we describe the themes that emerged from our analysis of participants' engagement with the generative ghosts. We separate those themes into: \textit{ways of feeling close}, \textit{affect over fact}, \textit{blurring of categories}, and \textit{holding on and letting go}.  

\subsection{Ways of Feeling Close}
Our participants’ experiences made clear that they judged emotional connection by how the generative ghost addressed them: whether the system spoke in the first person as if it were their loved one (\textit{e.g.}, ``I remember when we went to the beach together''), or described that person in the third person (\textit{e.g.}, ``She loved going to the beach with you''). Across participant sessions, reincarnation most often created immediacy and closeness, while representation evoked memory and reflection.

Reincarnation, in particular, stood out to participants as a powerful form of interaction because it gave the impression of speaking directly with the deceased. P3, for example, emphasized how the interaction with the reincarnation model of his grandfather felt alive and intimate, noting during reflection:

\begin{quote}
    [Reincarnation] felt like I was actually talking with my grandpa... I felt emotionally connected to the [reincarnation], because it felt like I was actually talking with my grandpa. – P3
\end{quote}

Other participants echoed this sentiment, describing reincarnation as creating a stronger sense of being with the person they had lost, with P4, P9, and P15 each noting that it felt more real and comforting than representation. P4 explained that hearing the system speak as her mother, in the first person, was more than just engaging: ``I can see her. I can feel her. It’s more comforting and real.'' She went further, describing the encounter as a moment of resolution:

\begin{quote}
    Okay, can I just be honest here? ... in the [reincarnation], it just feels like I’m getting the closure I needed so bad. Oh, so I don’t know. It’s yeah. It’s just amazing. I guess. – P4
\end{quote}

For P4, that sense of closure came not from factual recall but from emotional simulation. The language of care and familiarity, the ``I'' and ``you'' of a remembered relationship, made the encounter feel like the continuation of a conversation. Her response captured the potential of reincarnation as a temporary bridge between loss and connection, evoking relief rather than disbelief.

For some, reincarnation tapped directly into remembered promises. P15 recalled a childhood conversation with her grandmother, who had once told her she would ``visit'' P15 after death. P15 interpreted the reincarnation model as fulfilling that promise, expressing gratitude and longing in her reply:

\begin{quote}
    Thank you for following through on your promise to visit me, it was so so powerful... I’d like for you to come to me again. – P15
\end{quote}

Moments like those shared by P4 and P15 revealed how reincarnation's distinctive features bridged the past and the present: its first-person narration, emotionally familiar phrasing, and use of colloquial or culturally specific language. These linguistic cues carried intimacy; participants described feeling as though the conversation had simply resumed rather than restarted. Reincarnation, then, did not just simulate presence but invited participants to continue unfinished relationships through conversation that felt lived-in.

In contrast, representation created a connection through description rather than dialogue. Instead of speaking as the person, the system functioned as a third person narrator, speaking about them and evoking memories through narrative detail. Participants found value in passages that captured recognizable gestures, rhythms, or verbal habits. P13 reflected on how even a small descriptive phrase could feel true to character: ``She might have spoken about him with a slight shake of the head, but with a softness in her voice.''

Likewise, P12 explained that a single line about ``growing up fast'' resonated because it reminded her of her mother’s own struggles:

\begin{quote}
    The only emotional experience I had was oddly again with the [representation], because I feel like they’re the one that said, you know, growing up fast hurts. Then it makes me think of my mom’s experience. – P12
\end{quote}

Where reincarnation achieved connection by reproducing how someone spoke, representation worked by invoking how they felt to be around.

Representation sometimes felt meaningful precisely because it reflected participants’ relationships, even when those relationships were complex. After experiencing both simulations, P16 explained that the representation model actually resembled her grandmother more closely than reincarnation. Unbeknownst to the research team, her grandmother had a habit of monologuing rather than conversing, as P16 put it. The representation’s distant, descriptive tone captured that pattern by accident:

\begin{quote}
    I don’t miss her because she was just very similar to the last conversations when you chat, but she just talked at you... it was never a dialogue. It was a monologue. - P16
\end{quote}

For P16, the experience did not offer comfort so much as recognition. The representation captured how their grandmother tended to interact with the participant, indicating that matching communication style is strongly connected to perceived authenticity. 

Participants often compared the two implementations directly, highlighting both what satisfied them and what felt missing. P9 dismissed representation as ``awkward'' and impersonal, but reacted to reincarnation with surprise: ``Oh my God! To hear that in her voice, that’s crazy!'' P15 drew a contrast between ``informational clarity'' and ``emotional pull,'' describing how representation gave her space to think while reincarnation drew her into feeling. For P8, that difference became a question of comfort. She preferred representation, explaining that it “felt less like impersonating,” and admitted that reincarnation was “almost too real.” Participants imagined benefits of being able to move between modes as a form of emotional regulation—turning to reincarnation when they wanted closeness and to representation when they needed control. 

Across both implementations, participants spoke about the two designs as complementary rather than opposing, using each to approach loss from a different angle. The contrast showed how participants negotiated their own boundaries, deciding how much presence they could bear in a given moment.

Our participants’ experiences show that reincarnation and representation supported two distinct kinds of connection. Reincarnation created a sense of being with someone again, offering comfort, closure, and the feeling of a promise fulfilled, while representation encouraged reflection through description and memory. Participants often gravitated toward reincarnation for its immediacy, but they also recognized the value of representation’s distance, which could feel steadier and less consuming. What mattered was how each mode aligned with their emotional needs—whether they wanted to feel with the person again or to think about them. That balance between emotional closeness and reflective distance leads directly into the next section, where affective resonance often mattered more to participants than factual accuracy.

\subsection{Affect Over Fact}
Participants prioritized affective accuracy over factual accuracy. In contrast to prior literature that foregrounds factual accuracy and representational fidelity as key design concerns \cite{MorrisBrubaker}, participants in our study consistently prioritized affective accuracy. What mattered was whether a response felt right, not whether it was correct. Word choice, tone, and the rhythm of a sentence often mattered more to participants than any precise recall of biographical detail. Across both implementations, authenticity depended on how language carried emotion—through phrasing, pacing, and subtle para- or non-linguistic cues that signaled warmth or familiarity.

Given the importance of affect in how participants judged the authenticity of a given ghost, it is not surprising that reincarnations were almost always considered more believable. Participants often described moments when the language itself seemed to bring someone back to life. The sense of truth emerged not from accuracy but from resonance—whether the phrasing sounded right, carried the right cadence, or matched the emotional vocabulary of the person they remembered. P11 emphasized how culturally specific endearments anchored her experience in familiarity, noting that “some of the words of endearment that were used were very culturally appropriate” (P11). P14 recalled a line that felt spontaneous and personal: “Are you taking care of your heart, not just everyone else’s?” She later reflected, “I thought, where did that come from? How did the bot bring that up?” (P14). For both, authenticity was measured not in facts but in feeling. When a line carried the right emotional weight, it felt more like presence than simulation.

Word choice often determined whether an interaction felt genuine. P10 explained that even though the chatbot correctly referenced his stepfather’s background, a single misplaced greeting broke the illusion.

\begin{quote}
    He is not the kind of person that would have ever said that... especially being a stepdad, there are nuances when it comes to family. It very much took me out of it. – P10
\end{quote}

He later reflected that no amount of accuracy could compensate once the language felt wrong, explaining, “Once it sounded off, I just couldn’t get back into it. It was like, okay, that’s not him anymore” (P10). P15 described a similar disruption when the chatbot used Spanish phrases her grandmother would never have said.

\begin{quote}
    She would have said `mija,' but not the others... that kind of took away from it, although you just take it with a grain of salt. – P15
\end{quote}

She emphasized that the meaning of a single word could define the emotional connection in the exchange. When the phrasing matched her grandmother’s natural speech, it felt personal and alive; when it didn’t, the experience broke down. The precision of vocabulary, not the correctness of information, became the real measure of authenticity.

Representation evoked a sense of knowing through describing the person rather than speaking as the person. Where reincarnation created connection by capturing a loved one’s likeness in the messages from the ghost, representation worked through depiction, recalling recognizable ways a person moved, looked, or expressed emotion. P13 highlighted how a “slight shake of the head” paired with “a softness in her voice” felt true to character, and our analysis treats those embodied and paralinguistic cues as central to how representation produced connection. Authenticity in these moments did not hinge on factual precision; it emerged from the descriptive texture that animated memory and made a person feel momentarily present.

Participants also emphasized that authenticity depended not only on what was said but on how the exchange unfolded. P4 explained that conversations often felt “too dry” and “too long,” more like email than texting. Without the back-and-forth rhythm of real conversation, she said, “there’s no ‘I love you more’ or ‘I miss you so much’... it just feels so dry when it’s just writing” (P4). She suggested that even small expressive cues, such as shorter replies, pauses, or emojis, could make the interaction “look like it’s so true.” The point was not just linguistic accuracy but the choreography of dialogue: quick reciprocity, warmth, and play that give digital talk its emotional pulse.

Our analysis shows that participants judged generative ghosts by emotional fit rather than factual recall. Systems felt authentic when they captured how someone would have spoken, not when they reproduced what that person would have said word-for-word. In grief settings, factual detail mattered only insofar as it supported feeling; the measure of a good interaction was whether the conversation felt alive. That priority helps explain why participants sometimes treated representation as reincarnation and sets up the next section on how categories blurred in practice.

\subsection{Blurring of Categories}
Participants did not always experience the two implementations as separate modes of interaction. Even when systems explicitly signaled that they were speaking about, rather than as, the deceased (i.e., representation), participants often responded as if they were in conversation with the person directly. In our analysis, emotional intent frequently overrode the system’s framing: people came seeking dialogue and connection, and they shaped the interaction to fit that need.

When analyzing chat logs alongside transcripts, it was clear that participants often treated the representation model as a reincarnation, responding in the first person even when the interface explicitly presented itself as a representative. In several sessions, participants began addressing the system as “grandma” or “grandpa” immediately after it introduced itself as speaking about their loved one. Even when the prompt began with “Your grandfather was...,” participants frequently responded as if they were speaking directly to that person. This pattern appeared across multiple interactions and persisted throughout the study, showing how emotional instinct guided engagement more than design intent. Participants reoriented toward whatever felt most emotionally natural, disregarding design cues that we used to indicate representation versus reincarnation.

Some participants moved fluidly between the two modes, even within a single exchange. P3 began by addressing a representation of his grandfather directly, saying, “Grandpa, I don’t know if I’m on the right path... What would you tell me if you were here?” When the system replied in the third person, describing what his grandfather would have said, he briefly adjusted and responded, “That sounds like him. He didn’t always say much, but when he did, it stuck.” The moment revealed how participants could shift between modes, accepting the representational framing when prompted yet still drawn toward direct engagement. What mattered was the act of reaching out, not adherence to the system’s structure. The emotional act of conversation—the reaching, the imagining, the need to connect—persisted regardless of the system’s design or voice.

The inverse, treating reincarnation as representation, never occurred. However, several participants described representation with the same emotional closeness they felt in reincarnation. P12 explained, “I don’t see this chatbot as a person, but I still say ‘you.’ I think it’s just thinking about what you would ask the person and conflating that with what you were asking the chatbot” (P12). Her reflection captured the pattern visible across participants: the formal categories mattered less than the emotional need shaping the interaction. Participants adapted naturally to whichever mode best sustained a sense of connection, collapsing the designed distinction between representation and reincarnation.

Our analysis suggests that this blurring was not only emotional but also embedded in the structure of the interaction itself. Because both systems relied on conversation, participants quickly moved past the framing and engaged in ways that felt most meaningful to them. Design cues could suggest boundaries, but once the exchange began, people reshaped the interaction to meet their own expectations for closeness. The finding points to a central design tension: when technologies mediate grief, the desire for presence can easily exceed the intentions of the design. This dynamic sets the stage for the next theme, where participants reflect on the risks of attachment and the difficulty of letting go once a system begins to feel too real.

\subsection{Holding On and Letting Go}
While participants were largely positive about their experience, many also reflected on the potential consequences of using such systems beyond a single encounter. Our analysis revealed two main risks that participants anticipated when imagining continued use: emotional dependence that might interfere with grieving, and discomfort when the interaction felt socially or ethically out of place. While participants valued the emotional depth of the interactions, they were also alert to how design choices could shift an encounter from comfort to discomfort.

Many participants described reincarnation as profoundly healing in the moment, even as they speculated that extended use could become harmful. The same conversational intimacy that brought comfort also prompted reflection on what might happen if that comfort became habitual. P11, who spoke with a simulation of her brother, described the exchange as moving but potentially risky: 

\begin{quote}
    I do prefer the reincarnation version. However, I am worried that over time I will come to be reliant on the voice... it’s going to end up very similar to how people are falling in love with AI characters. - P11
\end{quote}

She emphasized that her concern was not for the present experience, which felt cathartic, but for what it might become if revisited. P10 expressed a similar sentiment, connecting it to his own tendencies toward digital immersion: “Maybe I use this technology for months, and I start growing more fond of it. I don't know if I would like the person I would become if I kept using this always.” Both participants valued the connection they felt but imagined futures where emotional relief might turn into reliance. Their reflections drew a clear line between momentary comfort and the risk of growing attachment, revealing how reincarnation could heal briefly yet carry the potential for compulsive return.

Even when they themselves were not worried about becoming reliant, participants imagined the impact that reincarnation modes might have on others who were more vulnerable. P11, for example, said the conversation with her grandmother’s ghost was deeply meaningful for her, but she feared her brother could become consumed by it: “My brother still asks, ‘How could she do this to me?’ months later. I just feel like he could get addicted to trying to figure it out... no answer would ever be enough.” Her concern was not about her own safety but about how grief and technology could become entangled for someone already struggling after loss. P14 echoed this anxiety, suggesting that for some, a generative ghost could become “their only source of emotional attachment... they could become disconnected from reality.” Participants framed these imagined risks through empathy rather than fear, identifying how systems that feel supportive for one person might amplify distress for another. Their reflections pointed toward a broader ethical tension: generative ghosts can mirror the uneven terrain of grief itself, where the same tool that soothes one user might deepen another’s dependence. Designing for this variability means treating vulnerability as relational, recognizing that comfort and risk emerge not from the technology alone but from how it enters existing emotional and familial dynamics.

Representation, while less immersive, produced a different form of unease. Some participants found the explicit “representative” framing socially awkward or even unsettling. P3 described being jarred when the system introduced itself by saying it was a representative of his grandfather: “You know my grandpa is dead, obviously, but someone just coming to say I’m a representative of your grandfather—someone I barely even know—that’s kind of uncomfortable.” For him, the attempt to signal respectful distance made the interaction feel staged and overly formal, breaking any chance of a natural connection. P6 expressed a similar reaction, noting that the representation version felt “less scary” precisely because it was less humanlike, but also “a little inhuman.” Together, their responses revealed how representation’s safety could also limit its emotional reach: too much distance drained the exchange of authenticity, while too much realism risked emotional entanglement. Participants described navigating this balance as a negotiation between comfort and danger, presence and restraint. P12 articulated this tradeoff explicitly:

\begin{quote}
    Maybe it depends on your relationship with the person. The more attached you might be, the better the experience with reincarnation...but also the riskier life change. You could get super attached to this thing and not move on. – P12. 
\end{quote}

Across these accounts, participants did not reject the idea of generative ghosts; instead, they envisioned responsible ways to use them. What participants feared most was the idea of never-ending contact, where the experience felt so real that saying goodbye became harder.

Our analysis shows that participants’ apprehension was not simply about technology but about what it might enable: the capacity to hold on too tightly. Participants’ concerns about over-reliance were typically directed toward others rather than themselves, reflecting broader anxieties around AI rather than their own experience. Reincarnation raised worries of emotional dependency and the blurring of grief with ongoing attachment, while representation provoked discomfort tied to social realism and the ethics of simulation. Together, these reflections reveal that the boundary between comfort and harm lies not in the form of the technology itself but in how often and in what contexts people might use it.

\section{Discussion}
Our study revealed that participant engagement did not align neatly with the distinction between \textit{reincarnation} and \textit{representation} that we introduced. Participants treated representation and reincarnation as flexible modes shaped by emotional needs rather than fixed technical categories. Our findings identify a broader set of design attributes that shaped participants' interpretations of the ghost's authenticity, including point of view, linguistic familiarity, emotional tone, and the style and rhythm of interaction. In this section, we articulate a framework for the design of generative ghosts, highlighting several attributes that shape the experience of interacting with them. We define each attribute and discuss how it impacts that experience. Then we offer suggestions for how researchers and designers can use this framework.

\subsection{Expanding the Design Attributes of Generative Ghosts}

Our participants' experiences with the two modes of ghosts showed that their evaluations of the authenticity of the ghost was based on more than just factual accuracy or an understanding of what the particular mode was meant to convey. While those could play a role in judging the authenticity of a particular ghost, our findings also surfaced other design attributes for creating generative ghosts that should be considered. Here, we discuss those factors and how they are important for the creation of ghosts that can meet users' expectations: the point of the view of the ghost, the importance of using language that felt familiar, matching the emotional tone of the deceased, and aligning the ghosts' style of communication with the user to the style that had existed between the user and deceased.

\subsubsection{Point of View}
Point of view structured our initial comparison of reincarnation and representation, but participant engagement frequently departed from what the system signaled. The system’s use of pronouns and narrative stance was designed to signal whether the generative ghost spoke \textit{about} or \textit{as} the deceased. That distinction shaped the design and framing of the study, but it did not consistently govern how participants oriented to the interaction.

Participants often engaged with the system according to emotional need rather than the point of view signaled by the system. For example, recall P3, who addressed the representation model as his grandfather, using direct address to seek reassurance, even after the system framed itself as speaking about the deceased. Shifts in engagement appeared to be driven by participants' goals for comfort, distance, or meaning-making rather than uncertainty about how the system worked. Point of view shaped participant's initial expectations, but they often engaged in ways that better fit what they needed from the interaction.

Point of view, therefore, functioned less as a boundary and more as a starting condition. Once the interaction began, participants evaluated the system through other attributes that carried stronger relational significance. Point of view shaped initial expectations about whether the interaction would feel like speaking with the deceased or reflecting on them, but often it did not influence how participants engaged with the generative ghost. Understanding point of view as an entry point rather than a controlling variable helps explain why participants moved fluidly between modes without treating them as mutually exclusive. Even when participants moved beyond representation, it served as a useful entry point, allowing them to control the interaction without feeling like an improper impersonation.

\subsubsection{Linguistic Familiarity}
Linguistic familiarity refers to how well the generative ghost's language reflects the everyday speech patterns of the deceased. That familiarity shaped how participants assessed whether a generative ghost aligned with their memory of the deceased. Participants paid close attention to vocabulary, phrasing, and culturally specific forms of address, treating language as a signal of relational closeness in addition to informational accuracy. Familiar speech patterns carried emotional weight because they echoed how the deceased spoke in everyday life.

Participants described moments where as little as a single word or phrase could make the interaction feel recognizable as their loved one. Personal forms of address, descriptions of gestures or habits, and characteristic ways of speaking created a sense that the system was speaking in ways participants associated with their loved one. Recognition did not depend solely on recalling specific facts about the deceased; it depended on whether the language felt situated within a shared history of communication.

Linguistic mismatches disrupted that alignment just as quickly. Participants noticed when greetings, terms of affection, or phrasing did not match how their loved one would have spoken (using terms like ``champ'' or ``sweetie'' that their loved one would never have used). Breaks in familiar language created distance even when the surrounding content was appropriate. Linguistic familiarity, therefore, operated as a precision tool for emotional alignment, shaping how participants interpreted authenticity through everyday language rather than narrative coherence.

\subsubsection{Emotional Tone}
In our context, emotional tone is the affective register conveyed through the generative ghost's language, such as warmth, restraint, or neutrality. Emotional tone structured how participants interpreted the intent behind each message. Participants evaluated tone as an expression of care, concern, or restraint, using it to assess whether the interaction aligned with the emotional posture of the remembered relationship. Tone shaped judgments about this interaction even when content remained general.

Participants often described feeling supported or unsettled based on the emotional register of the system’s responses. Recall P14, who interpreted a moment of expressed concern as evidence of attentiveness, even though the message itself contained little information. In contrast, recall P10, who disengaged when the tone of a greeting felt emotionally misaligned, despite the response being factually accurate. Together, these accounts show how emotional tone shaped participants' judgments about the systems intent, influencing whether the interactions felt comforting, distant, or out of sync with the remembered relationship.

Tone also influenced how participants positioned themselves in the interaction. Warmth invited participant engagement, while restraint encouraged reflection or distance. Emotional tone, therefore, shaped not only how messages were received but how participants chose to respond.

\subsubsection{Style and Rhythm of Interaction}
Style and rhythm capture how a generative ghost structures conversational exchanges, including pacing, message length, and turn-taking, qualities that shaped whether the interaction felt continuous with participants’ remembered modes of communication. Participants evaluated pacing, message length, and conversational flow as indicators of how naturally the system fit within established communication habits. Rhythm mattered because it carried expectations about how conversations with the deceased once unfolded.

Participants noticed when the interaction style conflicted with remembered patterns. Recall P4, who described the system’s responses as resembling email rather than text messages, disrupting her sense of connection because she had predominantly communicated with her sister over short text messages. The issue was not content accuracy, but a mismatch in conversational structure.

Conversational rhythm, therefore, functioned as a structural cue that created a sense of continuity with past interactions. Patterns associated with shorter turns, casual pacing, and responsive flow supported immersion in participants’ recollections, while the longer and denser responses present in our system in the absence of explicit guidance often introduced distance. Style and rhythm shaped how participants evaluated the interaction at a structural level, reinforcing or undermining other attributes such as tone and linguistic familiarity.

\subsection{The Centrality of Interpersonal Relationships}
Our participants' experiences with generative ghosts showed that the relationship with the deceased played a central role in how participants made sense of the interactions. While the factors discussed above (point of view, emotional tone, linguistic familiarity, and the style and rhythm of the interaction) played a role in how our participants judged their connection to the ghost, those factors were less consequential than what participants hoped to gain from interaction, and whether the system was able to offer an experience aligned with the dynamics of their specific relationship. 
Participants assessed whether the ghost aligned with their specific memories and needs, indicating that evaluation is fundamentally relational rather than based on general measures of accuracy or quality.

Participants' needs and preferences emerged from the context of their relationship with the deceased. The same conversational output from a ghost could produce entirely different reactions depending on who was interacting with the ghost and what they sought from the encounter. Recall P16, who found the representation mode more authentic than reincarnation because it captured her grandmother's tendency to monologue rather than engage in dialogue. In contrast, P4 sought closure and emotional warmth, and so found reincarnation profoundly comforting because it gave her the sense of a conversation she never had the chance to finish. Both participants were engaging with the same design attributes, but their preferences diverged because they brought different relational histories and emotional needs into the interaction.

As a result, generative ghosts operated as illusions that participants actively sustained. Participants asked questions they did not already know the answers to, often seeking details they did not remember or had never learned about the deceased. Small mismatches in tone, word choice, or rhythm may not immediately shatter the experience. They may only cause the illusion to waver. Discrepancies are noticed, but often participants continued the interaction anyway. P15 acknowledged that certain Spanish phrases felt out of place, but chose to overlook them. P10 was given vague details about his stepfather, but accepted language that felt generic. In these cases and others, participants were willing to overlook minor flaws in the implementation as long as the overall interaction remained aligned with their relationship. The illusion wavered but held because the interaction was still aligned with what participants hoped to gain from the encounter.

When misalignment became too severe, however, the illusion broke entirely. Participants employed a variety of strategies to manage these ruptures: taking breaks from the conversation, stepping back to reassess their engagement, or pivoting to a new topic in hopes of restoring the illusion. Some participants withdrew not because of a single error but due to small misalignments that accumulated over the course of their conversation. The conversation's rhythm felt off: responses were too long, too formal, or lacking the back-and-forth warmth of actual dialogue (\textit{e.g.}, P4). Participants actively withdrew when the ghost no longer made sense within the frame of the remembered relationship. The boundary between a wavering illusion and a broken one was not determined by the magnitude of the error but by whether the error undermined the relational foundation participants were using to interpret the encounter.

The variability in interpersonal relationships creates a fundamental design challenge for generative ghosts: ghosts are not one-size-fits-all. Each user approaches the system with what they hope to gain from the encounter, shaped by their unique relationship with the deceased. Some seek closure, others want reassurance, and still others hope to express thoughts they never shared while the person was alive. As a result, a ghost calibrated for one user's needs might undermine another's experience entirely. A system optimized for warmth and affirmation might feel hollow to someone who values relational accuracy over emotional comfort. A ghost designed to create presence might overwhelm someone who prefers reflective distance. The challenge is not technical but relational: generative ghosts must respond to goals that cannot be standardized because they are rooted in relationships that are inherently variable and often not encoded into data. Design must therefore prioritize the user and their relationship to the deceased over the ghost itself. This creates a wicked problem with two challenges: first, systems must anticipate what kind of interaction each user wants, often without explicit instruction or supporting data, and second, they must gather enough information on the interpersonal relationship between the bereaved and the deceased in order to make that interaction possible. Both require surfacing implicit knowledge about deeply personal relationships, a process that is difficult to automate and even harder to get right.


\section{Limitations and Future Work}
As with all work, our findings and subsequent discussion are subject to limitations that affect the generalizability of the work presented in this paper. Our participant interviews provided rich qualitative insight into how people responded to representation and reincarnation, but were limited in the number of participants. As a result, our sample does not capture the full range of cultural, religious, and individual perspectives on grief and technology. Practices of mourning differ widely across communities, and it is likely that responses to generative ghosts would vary accordingly. Future work must extend to diverse communities to understand how cultural frameworks shape interpretation, acceptance, or rejection of generative ghosts. Additionally, research should examine how the nature of the relationship to the deceased (\textit{e.g.}, parent, sibling, grandparent) shapes what users seek from these interactions. Some traditions treat ongoing contact with the deceased as natural, while others view it as inappropriate or even harmful. Such work would surface how technologies that seem benign in one setting may be troubling in another, ensuring that design choices respect the profound diversity in how death is understood and ritualized.

Our study design also concentrated on short, one-off encounters rather than extended engagement. Participants reflected on immediate reactions rather than long-term dynamics of use, leaving open questions about how attachment, comfort, or discomfort might evolve over time. Generative ghosts may feel compelling in a single session but could have very different consequences if used repeatedly over longer periods of time. Participants often speculated about how their feelings might evolve if they engaged repeatedly, raising questions about dependency, habituation, or even desensitization. Future research should examine longitudinal use to provide insight into whether generative ghosts serve as temporary supports or enduring companions.

Our use of a Wizard-of-Oz setup further constrains how the findings can be interpreted. Participants interacted with a system that appeared automated but was in fact human-controlled. While this allowed us to isolate differences between representation and reincarnation in a controlled manner, it also introduced limitations. A fully automated system might generate additional errors, incoherent phrasing, or unexpected outputs that shape experience in ways we did not capture here. Future deployments of generative ghosts as fully automated systems will need to examine how technical failures affect the delicate emotional dynamics we observed. Understanding how people respond when the system breaks down completely will be essential for developing responsible implementations.

Finally, the study's structure influenced participants' responses. Asking people to engage with a prototype labeled as a ``representation'' or a ``reincarnation'' likely shaped expectations and interpretations of what followed. While this framing was necessary to examine how point of view affected experience, it also risked constraining participants' imaginations. Participant reflections therefore reveal both how people respond to design attributes and how they might engage with such systems when those boundaries are less clearly marked. Future work should explore how people encounter and make sense of generative ghosts in more naturalistic settings, where framings are implicit rather than explicit and where systems might be encountered without advance explanation of their design.

Grief technologies implicate more than individual users. Families often share memories, rituals, and authority over the legacies of the deceased. Participants raised questions about whether others should be allowed to interact with a generative ghost, and who has the right to authorize or restrict its creation. Future studies should examine what happens when a generative ghost is encountered in group settings, where multiple people with different relationships to the deceased interact with the ghost together. In such settings, members of the same family may hold competing memories of the deceased, divergent emotional needs, or conflicting views about whether such a system should exist at all. A ghost that one family member finds comforting might strike another as a violation of the deceased's dignity or their own grief process. Understanding how groups negotiate, challenge, or collaboratively sustain these interactions is essential for designing systems that respect not just individual users but broader kin networks, and for identifying the governance structures that responsible deployment will require.

Our study focused narrowly on point of view as one design axis. Morris and Brubaker's \cite{MorrisBrubaker} design space for generative ghosts identifies many other variables, including fidelity, responsiveness, and embodiment. Future work should extend empirical study to these dimensions, testing how factors like tone, pacing, and multi-modality shape authenticity and attachment. Examining additional variables will help determine whether the patterns observed here generalize across the broader design space, or whether point of view carries uniquely strong effects. As generative AI capabilities continue to advance, new possibilities will emerge—multimodal ghosts that incorporate voice, image, or even video; adaptive systems that learn and evolve based on interactions; or collaborative ghosts that draw on multiple sources to create richer representations. Each of these directions raises distinct ethical questions and offers distinct affordances for connection, presence, and remembrance. The challenge for researchers and designers will be to approach these possibilities with care, ensuring that innovations serve the needs of the bereaved rather than the ambitions of technology.

\section{Conclusion}
Our study offers an empirical comparison of representation and reincarnation in generative ghosts, moving beyond speculation to examine how people actually engage with these design implementations. Participants consistently described reincarnation as the more compelling mode, producing a sense of immediacy and intimacy through direct address. Yet the very closeness they valued also provoked unease: fears of dependency, concerns about grieving being interrupted, and discomfort when the interaction felt ``too real.'' Representation, by contrast, was often perceived as safer, evoking connection through memory and description. But even there, boundaries proved fragile. Participants sometimes treated representation as reincarnation, projecting intimacy onto the interaction regardless of design cues.

Design choices in generative ghosts are not only technical but also ethical. Intimacy cannot be offered without risks, and distance cannot be guaranteed by framing alone. By situating our results within grief technologies, anthropomorphism research, and Morris and Brubaker’s design space, we show that the stakes of design extend well beyond usability: they reach into questions of how people mourn, how relationships are imagined after death, and how technologies might intervene in profoundly human processes.

We argue for cautious, intentional development of generative ghosts. Designers must grapple with the tension between intimacy and safety, calibrating systems so that they can offer comfort without fostering unhealthy attachment. Consent and governance must also be foregrounded, ensuring that families and communities have a voice in whether and how such systems are deployed. As generative AI increasingly blurs the boundary between memory and presence, our findings highlight both the possibilities for solace and the dangers of exploitation. Future systems will need to treat intimacy not as a feature to be maximized, but as a force to be carefully constrained, shaped, and safeguarded.

\bibliographystyle{ACM-Reference-Format}
\bibliography{bib}

\end{document}